\documentclass[aps,twocolumn,superscriptaddress,citeautoscript,footinbib,longbibliography]{revtex4-1}
\usepackage{graphicx}
\usepackage{multirow}
\usepackage{xcolor}
\usepackage{times}
\usepackage{amsmath,amsfonts}
\usepackage{bm}
\usepackage{bibunits}
\usepackage{tikz}
\usepackage{dsfont}
\usepackage{mathtools}
\usepackage{setspace}
\usepackage[normalem]{ulem}
\usepackage[caption=false]{subfig}

\usepackage[colorlinks=true,linkcolor=magenta,citecolor=magenta,hypertexnames=false]{hyperref}

\begin{document}
\title{Unsupervised learning of topological phase diagram using topological data analysis}
\author{Sungjoon \surname{Park}}
\affiliation{Center for Correlated Electron Systems, Institute for Basic Science, Seoul 08826, Korea}
\affiliation{Department of Physics and Astronomy, Seoul National University, Seoul 08826, Korea}
\affiliation{Center for Theoretical Physics (CTP), Seoul National University, Seoul 08826, Korea}

\author{Yoonseok \surname{Hwang}}
\affiliation{Center for Correlated Electron Systems, Institute for Basic Science, Seoul 08826, Korea}
\affiliation{Department of Physics and Astronomy, Seoul National University, Seoul 08826, Korea}
\affiliation{Center for Theoretical Physics (CTP), Seoul National University, Seoul 08826, Korea}

\author{Bohm-Jung Yang}
\email[Electronic address:$~~$]{bjyang@snu.ac.kr}
\affiliation{Center for Correlated Electron Systems, Institute for Basic Science, Seoul 08826, Korea}
\affiliation{Department of Physics and Astronomy, Seoul National University, Seoul 08826, Korea}
\affiliation{Center for Theoretical Physics (CTP), Seoul National University, Seoul 08826, Korea}

\date{\today}
\let\oldaddcontentsline\addcontentsline
\renewcommand{\addcontentsline}[3]{}

\begin{abstract}
Topology and machine learning are two actively researched topics not only in condensed matter physics, but also in data science.
Here, we propose the use of topological data analysis in unsupervised learning of the topological phase diagrams.
This is possible because the quantum distance can capture the shape of the space formed by the Bloch wavefunctions as we sweep over the Brillouin zone.
Therefore, if we minimize the volume of the space formed by the wavefunction through a continuous deformation, the wavefunctions will end up forming distinct spaces which depend on the topology of the wavefunctions.
Combining this observation with the topological data analysis, which provides tools such as the persistence diagram to capture the topology of the space formed by the wavefunctions, we can cluster together Hamiltonians that give rise to similar persistence diagrams after the deformation.
By examining these clusters as well as representative persistence diagrams in the clusters, we can draw the phase diagram as well as distinguish between topologically trivial and nontrivial phases.
Our proposal to minimize the volume can be interpreted as finding geodesics in 1D Brillouin zone, and minimal surfaces in 2D and higher-dimensional Brillouin zones.
Using this interpretation, we can guarantee the convergence of the minimization under certain conditions, which is an outstanding feature of our algorithm.
We demonstrate the working principles of our machine learning algorithm using various models.
\end{abstract}

\maketitle

\section{Introduction}
Topology has been extremely fruitful in physics, and there have been tremendous efforts to exhaustively categorize the topological phases~\cite{kitaev2009periodic,hasan2010colloquium,kruthoff2017topological,bradlyn2017topological,po2017symmetry,shiozaki2018atiyah,watanabe2018structure,elcoro2020magnetic}.
Concurrently, topology has also become useful in data science.
In topological data analysis (TDA) \cite{edelsbrunner2000topological,zomorodian2005computing,carlsson2009topology,edelsbrunner2010computational,chazal2017introduction}, a rapidly growing field, the objective is to use data points to study the topology of the space from which the data was sampled.
This is done by building a continuous shape over the data, such as a simplicial complex. Then, one computes the persistence diagram (PD), which encodes the topology of the continuous shape, and therefore, the topology of the data distribution.
Incorporating TDA into machine learning is an important research topic.

Recently, there were various efforts to apply unsupervised machine learning to find the topological phase diagram \cite{rodriguez2019identifying,fukushima2019featuring,balabanov2020unsupervised,greplova2020unsupervised,che2020topological,scheurer2020unsupervised,long2020unsupervised,kaming2021unsupervised,tsai2021deep}.
Most of the proposed techniques involve a clustering algorithm using some notion of similarity between two phases that somehow captures their topology.
In particular, Ref.~\cite{rodriguez2019identifying} showed that spectral clustering methods such as the diffusion map~\cite{coifman2005geometric,nadler2005diffusion,coifman2006diffusion} can be used to classify topological phases.
This idea was further developed in the context of topological insulators in Refs.~\cite{che2020topological,scheurer2020unsupervised,long2020unsupervised} with various notions of similarity between phases.

\begin{figure}[t]
\centering
\includegraphics[width=8.5cm]{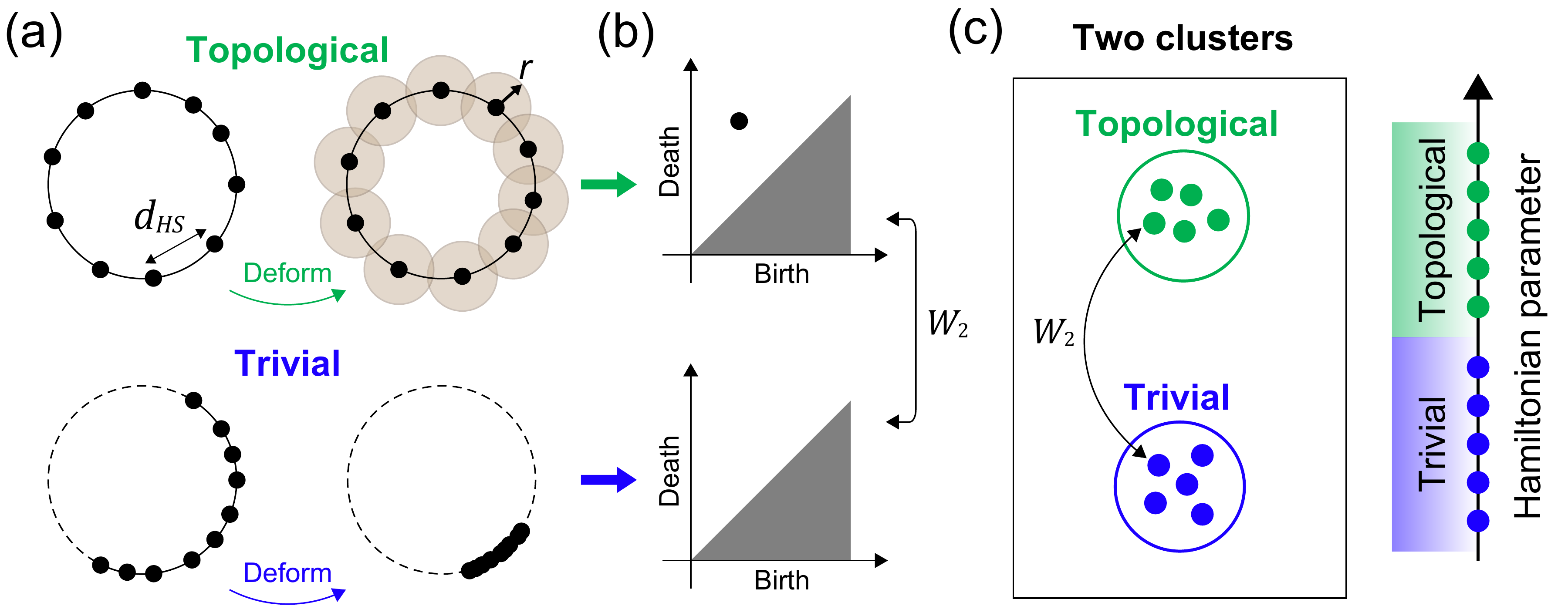}
\caption{(a) Embeddings of the wavefunctions sampled from the 1D Brillouin zone into $S^1\subset \mathbb{R}^2$, while preserving the Hilbert-Schmidt distance ($d_{HS}$). 
It is generally necessary to deform the wavefunctions to reduce the length of the space they span before computing the persistence diagram (PD).
The difference between a circle and an arc is captured by the homology of the union of balls of radius $r$ about the wavefunctions, whose changes are recorded in the PD shown in (b).
Birth (death) is the $r$ at which homology appears (disappears). The gray regions in PD is forbidden since death $>$ birth.
(c) By sampling the Hamiltonian for various parameters, we can compute the PD for each Hamiltonian as in (a) and (b), represented in blue and green dots.
Using the Wasserstein distance ($W_2$) between PDs, we can cluster similar PDs to obtain the topological phase diagram.
}
\label{fig.algorithm}
\end{figure}

In this work, we propose an unsupervised learning algorithm of topological phase diagrams that is based on TDA.
A bird's eye view of our approach is shown in Fig.~\ref{fig.algorithm}.
The motivation for the algorithm is that the quantum distance between wavefunctions such as the Hilbert-Schmidt distance ($d_{HS}$) \cite{buvzek1996quantum,dodonov2000hilbert} can capture the topology of the space formed by the wavefunctions, since isometry (distance preserving map) is a homeomorphism (topology preserving map).
For example, if the space formed by the wavefunctions over the one-dimensional Brillouin zone (BZ) forms a circle, or only part of a circle, the topological difference between them is encoded in $d_{HS}$, see Fig.~\ref{fig.algorithm}(a).

To draw the phase diagram using the above observation, we deform the wavefunctions to reduce the volume they span.
For example, suppose that the wavefunctions are constrained to live on a circle.
If the wavefunctions over the one-dimensional BZ forms a circle, this space will remain a circle even if we try to reduce the length of the space formed by the wavefunction.
However, if the wavefunctions span an arc, it will contract to a short arc (ideally a point), see Fig.~\ref{fig.algorithm}(a).

We then compute the PDs (see Fig.~\ref{fig.algorithm}(b)) by considering the union of the balls of radius $r$ (light brown disks in Fig.~\ref{fig.algorithm}(a)) about the data points (wavefunctions shown in black dots).
In a PD, we indicate how the topology of this union of balls changes as we vary $r$~\footnote{Note that here, we are considering the \v{C}ech complex.}.
In Fig.~\ref{fig.algorithm}(b), the $x$-axis (birth) is the $r$ at which the first homology becomes nontrivial, and the $y$-axis (death) is the $r$ at which the first homology becomes trivial.
Since the first homology of a circle and an arc over integer coefficient is $\mathbb{Z}$ and $0$ respectively, the PD is able to capture this difference.
By defining a similarity measure between PDs, we can cluster Hamiltonians whose wavefunctions yield similar PDs, and therefore, learn the topological phase diagram in an unsupervised manner as in Fig.~\ref{fig.algorithm}(c).

\section{Illustration with the SSH model}
Let us walk through our algorithm using the well-known Su-Schrieffer-Heeger (SSH) model~\cite{su1980soliton}.
The SSH model has the $\mathcal{PT}$ symmetry, where $\mathcal{P}$ is the inversion symmetry and $\mathcal{T}$ is the time-reversal symmetry.
If we choose the basis where $\mathcal{PT}$ is equal to the complex conjugation $K$, the Hamiltonian only has real components:
$H_{\bm{k}}=\sigma_x(t_1+t_2 \cos k)+\sigma_z t_2 \sin k$.
It is topological (trivial) when $|t_1| < |t_2|$ ($|t_1| > |t_2|$) because of the $\pi$-Berry phase ($0$-Berry phase).

\begin{figure}[t]
\centering
\includegraphics[width=8.5cm]{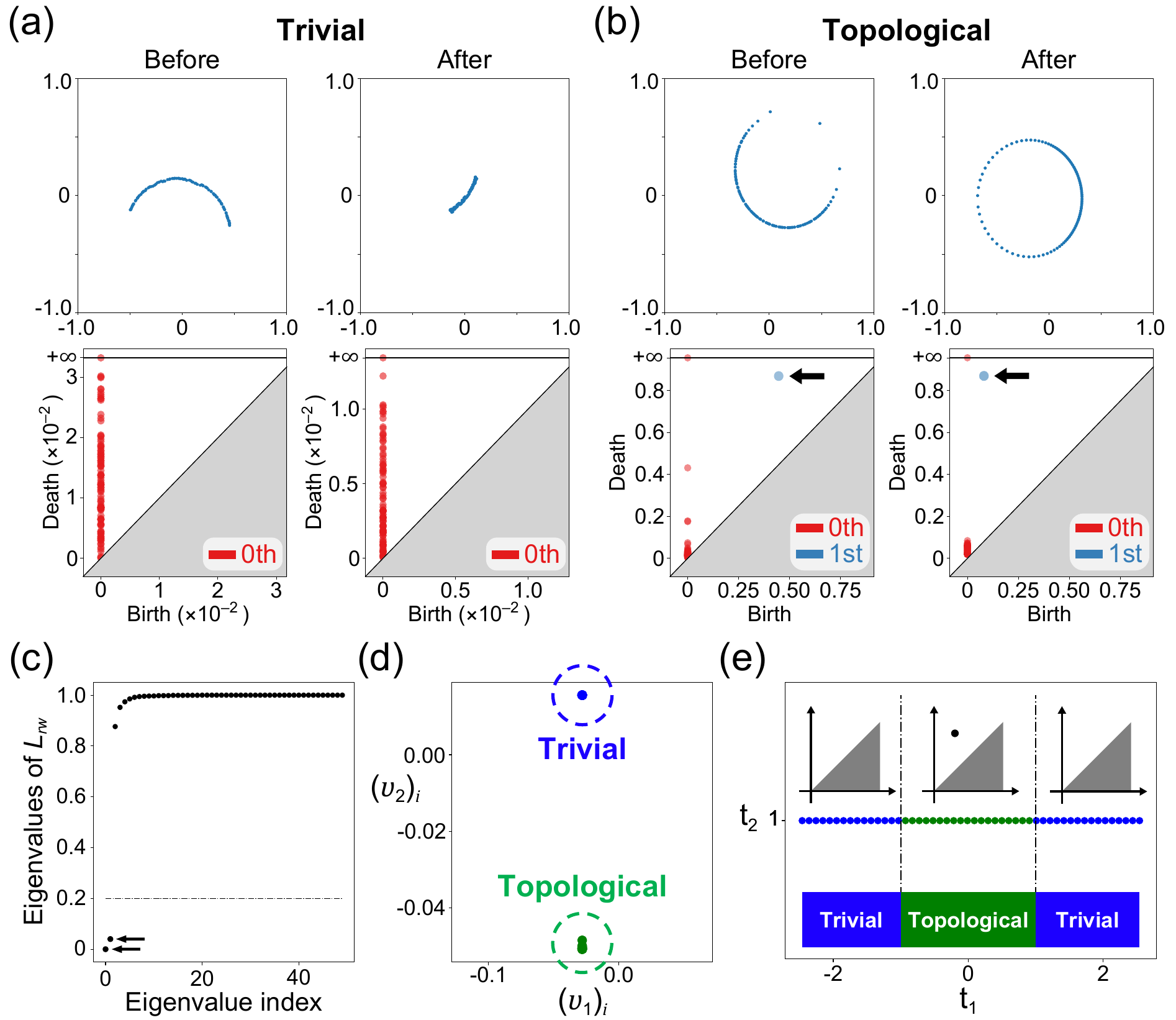}
\caption{(a) Topologically trivial SSH model, with $t_1 = 1$ and $t_2 = 0.95$. 
Left: mMDS and the PD before the deformation.
The legend in the PD indicates the color of the $n$th homology.
Only the zeroth homology appear in the PD (red dots correspond to the changes in the zeroth homology, see Appendix~\ref{app.tda}.
Right: mMDS and the PD after the deformation. Note the contraction of the space formed by the wavefunction.
(b) Topologically nontrivial SSH model, with $t_1 = 0.95$ and $t_2 = 1$. 
Left: mMDS and the PD before the deformation. 
Because of the regions with sparse wavefunction distribution, the first homology (indicated with black arrow) is hard to capture.
Right: mMDS and the PD after the deformation. Wavefunctions are more evenly distributed and the first homology is more easily captured.
(c) The spectrum of random-walk Laplacian ($L_{\textrm{rw}}$) formed using the similarity in Eq.~\eqref{eq.diagram_similarity}. 
We take the number of eigenvalues below $0.2$ to be the number of clusters.
We see that there are two clusters, indicated with black arrows.
(d) Plot of the components of the eigenvectors corresponding to the two smallest eigenvalues of $L_{\textrm{rw}}$.
The k-means clustering captures two clusters, which is used to draw the phase diagram in (e).
We can distinguish topologically trivial and nontrivial regions by examining the PD.
}
\label{fig.ssh_model}
\end{figure}

Because of the $\mathcal{PT}$ symmetry, the wavefunctions of this system live in the real projective space $\mathbb{R}P^1$, which is a circle.
The distance between two wavefunctions $\psi_1$, $\psi_2$ is given by the Hilbert-Schmidt distance:
\begin{align}
d_{HS}(\psi_1,\psi_2)=\sqrt{1-|\langle \psi_1 | \psi_2 \rangle|^2}.
\end{align}
Note that generally, real-valued wavefunctions (i.e. $\mathbb{R}P^{n-1}$) of $n$-band Hamiltonians provided with $d_{HS}$ can be isometrically embedded into the sphere $S^{\frac{n^2+n}{2}-2}$ in the Euclidean space $\mathbb{R}^{\frac{n^2+n}{2}-1}$~\cite{hioe1981n,jakobczyk2001geometry,kimura2003bloch,byrd2003characterization}.
When $n=2$, the wavefunctions can be embedded into $S^{1}$, so that we can easily visualize the space formed by the wavefunction using the metric multidimensional scaling (mMDS) \cite{borg2005modern}, which is an unsupervised learning algorithm that tries to embed the data points isometrically into the Euclidean space of a given dimension.
(See Appendix~\ref{app.dim_red}.)
We show the result of mMDS for the topologically trivial and nontrivial cases for the SSH model in the upper panels of Figs.~\ref{fig.ssh_model}(a) and \ref{fig.ssh_model}(b).
This is drawn by preserving $d_{HS}$ between the wavefunctions of the lowest energy band sampled at momentum $k_i$, where $\{k_i=\tfrac{2\pi i}{N_{\textrm{mesh}}}|i=0,...,N_{\textrm{mesh}}-1\}$, and $N_{\textrm{mesh}}=100$.

For the SSH model, the two spaces formed by the wavefunctions are already topologically distinct, and TDA can already capture this difference at this point, as was also observed in Ref. \cite{leykam2021photonic}.
However, a deformation process is necessary in more complicated models.
For $\mathcal{PT}$ symmetric $2\times 2$ Hamiltonian, continuous deformations can be written as $H_{k}(\theta_k) = U(\theta_k) H_{k}U(\theta_k)^T$, where $U(\theta_k)$ is a continuous function from $k$ to $SO(2)$ and $\theta_k$ is the parameter that characterizes $SO(2)$ matrices.
Note that this has the effect of transforming $\psi_k \rightarrow U(\theta_k)\psi_k$.

We deform the wavefunctions by minimizing the following loss function:
\begin{align}
L_{\textrm{tot}} (\Theta)= L_{v}(\Theta)+L_{s}(\Theta)+L_{c}(\Theta),\label{eq.loss_ssh}
\end{align}
where $\Theta = (\theta_1,...,\theta_{N_{\textrm{mesh}}})$ and  each of the individual loss functions is defined below.
We define $L_v = \sum_{i=1}^{N_{\textrm{mesh}}} d_{HS}(\psi_i,\psi_{i+1})$, where $\psi_{N_{\textrm{mesh}}+1} = \psi_{1}$.
Note that this a discretized approximation of the length swept by the wavefunction.
Minimizing $L_v$ can be interpreted as deforming the Hamiltonian to minimize the length of the space swept by the wavefunctions.
We define $L_{s}=r_s \sum_{i=1}^{N_{\textrm{mesh}}} d_{HS}(\psi_i,\psi_{i+1})^2$, where the parameter $r_s$ controls the ratio between $L_v$ and $L_s$.
$L_{s}$ has the effect of penalizing sparse distribution of wavefunction embedded according to $d_{HS}$, while also favoring a small length.
Finally, $L_c = r_c\sum_{i=1}^{N}|\theta_{i+1}-\theta_{i}| e^{-d_{HS}(\psi_i,\psi_{i+1})/\overline{d}_{HS}}$, where $\overline{d}_{HS}$ is the average $d_{HS}$ between the wavefunctions and $r_c$ controls the weight given to the continuity.
Notice that large variation between $\theta_i$ and $\theta_{i+1}$ is allowed if $d_{HS}(\psi_i,\psi_{i+1})$ is large, and vice versa.

We compare the result of mMDS before and after minimizing $L_{\textrm{tot}}(\Theta)$ in Figs.~\ref{fig.ssh_model}(a) and \ref{fig.ssh_model}(b).
Notice that $L_{v}$ decreases in the topologically trivial case, as intended.
However, $L_{v}$ does not change much in the topologically nontrivial case because $L_v$ is already close to $\pi$ before the deformation, which is the minimum length when $N_{\rm mesh} \to \infty$, see Appendix~\ref{app.min_vol}.
However, notice that regions which originally had sparse distribution of wavefunctions become more densely populated after the deformation.
This is useful because it is difficult to capture the topology of the space underlying the data with PD if there are sparsely populated regions.
Indeed, we see that the first homology is captured much earlier on in the PD~\footnote{Here, we have used the Vietoris-Rips complex for the computation.} if we perform the deformation, see the lower panels of Fig.~\ref{fig.ssh_model}(b).

Now, we use these results to obtain the phase diagram of the SSH model.
First, we fix $t_2=1$ and uniformly sample 50 values for $t_1$ between $-2.5$ and $2.5$, and compute the PDs following the above procedure.
A PD can be considered a collection of points.
Thus, we can define a distance measure between two PDs by using the distance between collection of points.
The Wasserstein distance is a popular choice, which we adopt for our work.
The concept of PD and the Wasserstein distance are reviewed in Appendix~\ref{app.tda}.
Let us use the Wasserstein distance $W_2$ between two PDs denoted by $\textrm{PD}_i$ and $\textrm{PD}_j$ ($i,j=1,\dots,50$) to define a similarity measure between them:
\begin{align}
w_{ij} = \exp(-W_2(\textrm{PD}_i,\textrm{PD}_j)/\overline{W}_2), \label{eq.diagram_similarity}
\end{align}
where $\overline{W}_{2}$ is the average $W_2$ between the PDs. 

Once we obtain the similarity matrix between the PDs, we can carry out the spectral clustering, which is an unsupervised learning algorithm (see Ref.~\cite{von2007tutorial}).
Following Ref.~\cite{von2007tutorial}, we use the random-walk Laplacian $L_{\textrm{rw}}=1-D^{-1}W$. 
Here, $W$ is a matrix with $(W)_{ij}=w_{ij}$ and $D$ is a diagonal matrix with $(D)_{ii}=\sum_{j} w_{ij}$.
The magnitudes of the eigenvalues of $L_{\textrm{rw}}$ indicate how fast the eigenstates evolve during the random-walk process defined by $L_{\textrm{rw}}$.
Because a state localized in a cluster evolves slowly in this random-walk process, the two small eigenvalues of $L_{\textrm{rw}}$ shown in Fig.~\ref{fig.ssh_model}(c)  implies that there are two clusters \cite{von2007tutorial}.
To determine which PD belongs to which cluster, we plot $((v_1)_i,(v_2)_i)$ in Fig.~\ref{fig.ssh_model} (d), where $v_{1,2}$ are eigenvectors corresponding to the two smallest eigenvalues.
This plot can be interpreted as the embedding of the PDs according to their similarities.
The two clusters of PDs, which represent topologically distinct phases, can then be captured using the k-means clustering.
Because each $((v_1)_i,(v_2)_i)$ corresponds to a PD of a Hamiltonian, we can use the clusters to draw the phase diagram as in Fig.~\ref{fig.ssh_model} (e).
Furthermore, let us note that for a topologically trivial phase, the point clouds forming the deformed space should ideally retract to a single point.
Thus, we should expect that the points appearing in the PD that persists for a long interval in the birth and death space.
That is, all points on the PD should lie close to the line birth = death.
Using this property, we can easily
distinguish between topologically trivial and nontrivial phases by examining the PD.
For the SSH model, we see that the point corresponding to the first homology in the PD lies far from birth=death line in the topological phase, while the trivial case does not give birth to the nontrivial first homology.

\section{Three-band model}
\begin{figure}[t]
\centering
\includegraphics[width=8.5cm]{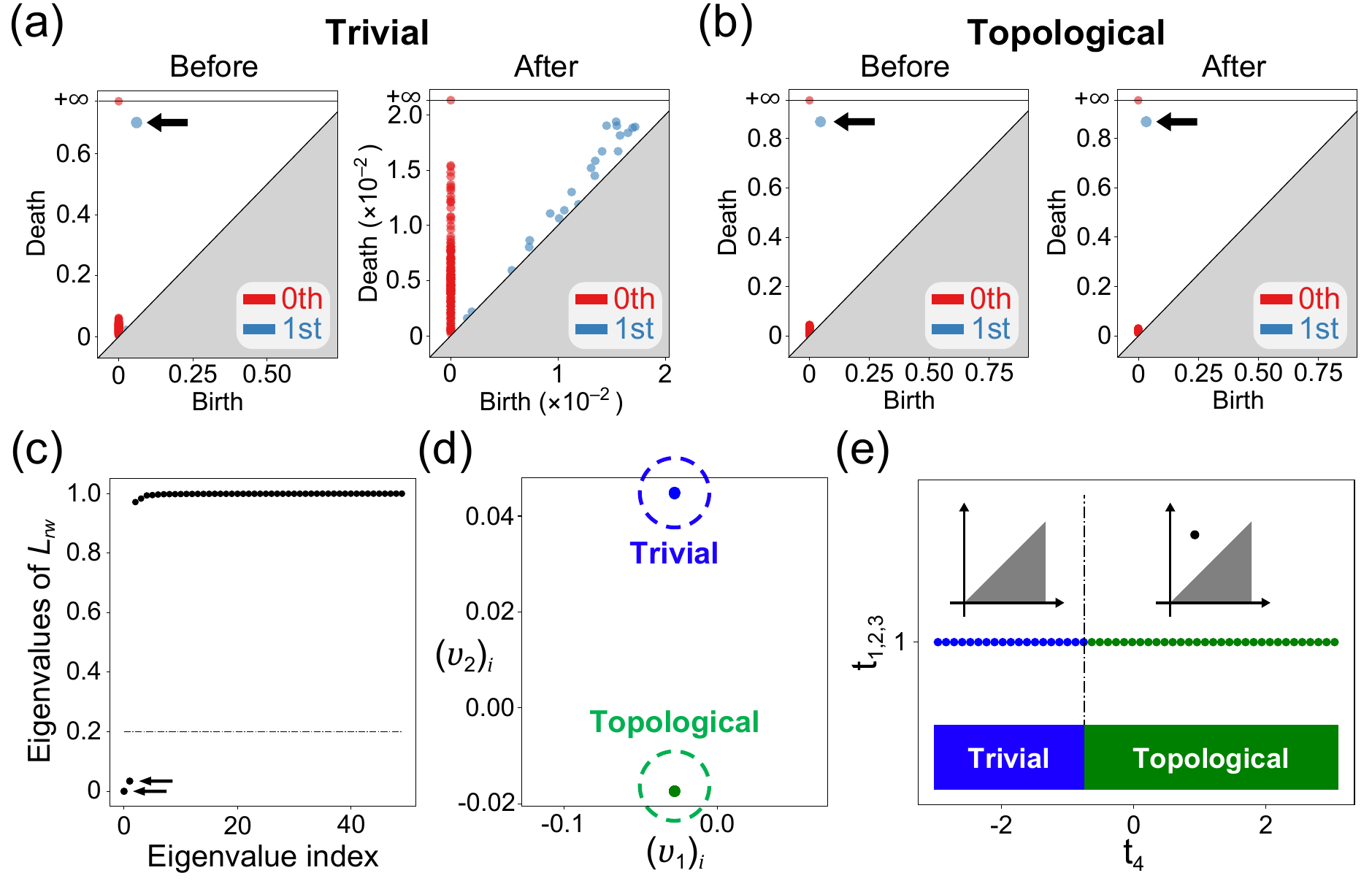}
\caption{(a), (b) Changes induced by the deformation in the PDs for the topologically trivial and nontrivial phases of the three-band model in Eq.~\eqref{fig.tb_model}, respectively. 
(c), (d) The eigenvalues and the eigenvectors of $L_{\textrm{rw}}$.
(e) The phase diagram is obtained using the k-means clustering.
}
\label{fig.tb_model}
\end{figure}

To demonstrate our algorithm in the case when there are more than two bands, we examine the following three-band Hamiltonian with $\mathcal{PT}=K$ symmetry (see Appendix~\ref{app.three} for details):
\begin{align}
H_{\bm{k}}=\begin{pmatrix}
t_2 \sin k & -t_1-t_2 \cos k & t_3 \\
-t_1-t_2 \cos k & -t_2 \sin k & -t_3 \\
t_3 & -t_3 & 2 t_4 \cos k
\end{pmatrix},
\end{align}
where $t_1, t_2, t_3, t_4$ are the Hamiltonian parameters.
For simplicity, we put $t_1=t_2=t_3=1$ and vary $t_4$.
The lowest energy band becomes topologically trivial (nontrivial) for $t_4<-0.75$ ($t_4>-0.75$) because of the $0$-Berry phase ($\pi$-Berry phase).
Because there are three energy bands, the states can be embedded isometrically in $\mathbb{R}^5$, so that it is difficult to accurately visualize the space formed by the states with mMDS, although other dimensionality reduction techniques can be helpful as explained in Appendix~\ref{app.dim_red}.
Instead, we examine the PDs computed with the lowest energy band, shown in Figs.~\ref{fig.tb_model}(a) and \ref{fig.tb_model}(b) with $N_{\textrm{mesh}}=150$, computed for $t_4 = -1$ and $t_4=-0.5$, respectively.
Notice that the deformation is critical for capturing the topology, since even in the topologically trivial phase (Fig.~\ref{fig.tb_model}(a)), the first homology before the deformation births at small $r \sim 0.06$ and dies only after $r$ reaches around $0.7$.
However, after the deformation, there is no first homology that persists for a large interval.
In contrast, in the topologically nontrivial phase (Fig.~\ref{fig.tb_model}(b)), the first homology persists for a large interval regardless of the deformation.

We obtain the phase diagram by sampling 50 points for $t_4$ between $-2.5$ to $2.5$ and proceeding as in the SSH model. 
From the eigenvalues of $L_{\textrm{rw}}$ in Fig.~\ref{fig.tb_model}(c), we see that there are two clusters, which can be confirmed by examining the corresponding eigenvectors in Fig.~\ref{fig.tb_model}(d).
In Fig.~\ref{fig.tb_model}(e), we show the phase diagram thus obtained, which is in excellent agreement with the one we computed using Berry phase.

\section{QWZ model}
Next, we demonstrate our algorithm in 2D with the Qi-Wu-Zhang (QWZ) model~\cite{qi2006topological}, whose Hamiltonian is
$H_{\bm{k}} = \bm{d}_{\bm{k}}\cdot \bm{\sigma}$, where $d^x_{\bm{k}}=\sin k_x$, $d^y_{\bm{k}} = \sin k_y$, $d^z_{\bm{k}} = \mu-2 b (2-\cos k_x - \cos k_y)$.
Here, the Hilbert space is $\mathbb{C}P^1$, which is a sphere. 
Note that generally, complex-valued wavefunctions ($\mathbb{C}P^{n-1}$) of $n$-band Hamiltonians can be isometrically embedded into $S^{n^2-2}\subset \mathbb{R}^{n^2-1}$, and for $n=2$, this is $S^2$.
(See Appendix~\ref{app.isometric}.)
The phase diagram of the QWZ model is as follows ($C$ is the Chern number): $C=0$ when $\mu/b < 0$; $C=-1$ when $0 < \mu/b < 4$; $C=1$ when $4 < \mu/b < 8$; $C=0$ when $8<\mu/b$.
For simplicity, we put $b=1$.

We can proceed as in 1D, except that the loss function needs to be slightly modified since we are working in 2D.
To do this, note that given the distances between three points, the area of the triangle $\Delta$ they form is given by $v(\Delta)=\frac{1}{4} \sqrt{2d_1^2d_2^2 + 2d_1^2 d_3^2 + 2 d_2^2 d_3^2 - (d_1^4+d_2^4+d_3^4)}$, where $d_1, d_2, d_3$ are the lengths of the three edges of $\Delta$.
Let us first divide the $k$ space along $x$ and $y$ directions into $N_{\textrm{mesh}}$ pieces, or equivalently, into $N_{\textrm{mesh}}^2$ squares.
Each square can be divided into two triangles, and we let $\{\Delta\}$ be the set of  $2N_{\textrm{mesh}}^2$ triangles thus obtained, whose edge lengths are given by $d_{HS}$ between the wavefunctions of the lowest energy band computed at the vertices.
We then define $L_v = \sum_{\{ \Delta\}} v(\Delta)$, which is the discretized approximation of the area swept by the wavefunction.
As before, we define the sparsity loss $L_s = r_s\sum_{\{ \Delta\}} v(\Delta)^2$ and the continuity loss $L_c = r_c\tfrac{1}{N_{\textrm{mesh}}}\sum_{\langle ij \rangle}(|\theta_{i}-\theta_{j}|+|\phi_{i}-\phi_{j}|+|\rho_{i}-\rho_{j}|) e^{-d_{HS}(\psi_i,\psi_{j})/\overline{d}_{HS}}$, where $\theta,\rho,\phi$ parametrizes the $SU(2)$ matrix that is used to deform the Hamiltonian.
Note that the overall ratio $1/N_{\textrm{mesh}}$ in $L_c$ is based on the observation that given a function varying only in one direction (say $k_x$), the summation increases linearly with $N_{\textrm{mesh}}$ for large $N_{\textrm{mesh}}$, in contrast to $L_v$, which converges to a constant.

Before proceeding as in 1D, it is useful to know that the computation time of gradient descent roughly increases linearly with the number of points in the $k$ mesh, while the computation time of PD increases drastically with the number of points in the $k$ mesh.
There are various methods that overcome this \cite{sheehy2013linear,buchet2016efficient,cavanna2015geometric}.
Here, we simply choose a large $N_{\textrm{mesh}}=40$ and sparsify the data by removing points that are too close to each other after the deformation until $240$ points are left.
Our method then becomes scalable to 2D.

\begin{figure}[b]
\centering
\includegraphics[width=8.5cm]{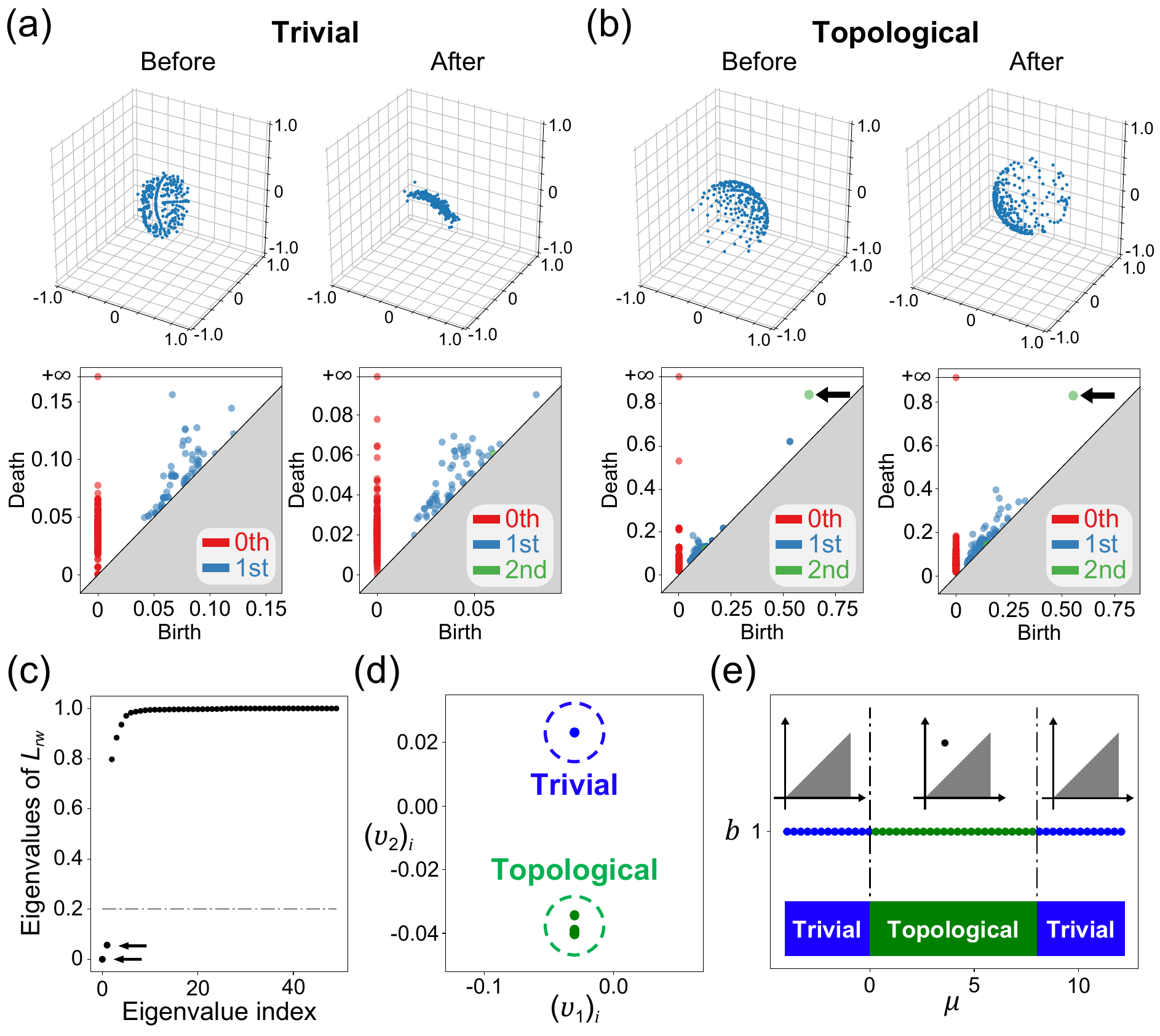}
\caption{(a), (b) Changes induced by the deformation in mMDS and PD for trivial ($\mu = -0.1$, $b=1$) and topological ($\mu = 0.1$, $b=1$) phases of the QWZ model respectively. 
The second homology is indicated with black arrows.
(c), (d) The eigenvalues and the eigenvectors of the random-walk Laplacian. Two clusters are captured by the k-means clustering, the result of which is used to obtain the phase diagram in (e).
We can distinguish topologically trivial and nontrivial phases with the second homology in the PD.
The topological region in (e) contains two phases with $C=-1$ and 1 respectively, which can be distinguished by using oriented area as detailed in Appendix~\ref{app.oriented_area}.
}
\label{fig.qwz_phase_diagram}
\end{figure}

We show the mMDS and PD for topologically trivial and nontrivial cases in Figs.~\ref{fig.qwz_phase_diagram}(a) and \ref{fig.qwz_phase_diagram}(b).
Notice that the wavefunctions cover a part of $S^2$ (all of $S^2$) in the topologically trivial (nontrivial) phase.
This difference can be captured using the second homology.

We obtain the phase diagram by fixing $b=1$ and sampling $50$ points for $\mu$ between $-4$ to $12$, followed by the clustering process, see Figs.~\ref{fig.qwz_phase_diagram}(c)-(e).
Notice that our method correctly distinguishes topologically trivial and nontrivial phases, and also provides information on which phases are topological through the PD.
We note that although the PD cannot distinguish the sign and sizes of Chern number because details of the map from BZ to $S^2$ is lost in the PD, we improve this limitation by introducing the oriented area to distinguish different Chern numbers (see Appendix~\ref{app.oriented_area}.)

\section{Discussion}
We showed that TDA combined with wavefunction geometry yields a machine learning algorithm that can distinguish between topologically trivial and nontrivial phases.
An important part of the algorithm is the minimization of the volume of the space formed by the wavefunction, which has the nice interpretation of finding geodesics (1D) and minimal surfaces (2D and higher). 
This interpretation allows us to analyze whether our algorithm will work.
As discussed in the Appendix~\ref{app.min_vol}, for 1D system with $\mathcal{PT}$ symmetry, we do not expect to become stuck at local minima of the length functional if we stay in the space of smooth curves in $\mathbb{R}P^n$;
similarly for 2D, we do not expect to become stuck at local minima of the area functional if we stay within the space of smooth submanifolds of $\mathbb{C}P^n$.
This theoretical guarantee on the convergence is an advantage of our algorithm, although, the computational cost (memory and time) can be large for complicated systems.

Although we have focused on the topology of a single band, our algorithm can be extended to multiband systems, since $d_{HS}$ can be defined for the occupied bands in multiband systems as well.
We can also extend our algorithm to symmetries other than $\mathcal{PT}$ symmetries. 
However, obtaining guarantees on the performance in these extensions is an open problem that we leave for future studies.
It would also be interesting to expand the analysis to fragile topology~\cite{po2018fragile} and non-Hermitian topology~\cite{gong2018topological,shen2018topological,yao2018edge}, as our approach gives a fresh perspective on the characteristics of topological systems.
Finally, it would be interesting to study the evolution of the wavefunction space across topological phase transitions.

\begin{acknowledgments}
SP thanks JHL for useful discussion.
During the preparation of our manuscript, we came across Ref.~\cite{leykam2021photonic}. Here, the authors observed that TDA can be used to capture the topology of the SSH model. 
However, they do not introduce deformation process, which is in general necessary to capture the topology.
S.P, Y.H., B.J.Y. were supported by the Institute for Basic Science in Korea (Grant No. IBS-R009-D1), 
Samsung  Science and Technology Foundation under Project Number SSTF-BA2002-06,
the National Research Foundation of Korea (NRF) grant funded by the Korea government (MSIT) (No.2021R1A2C4002773, and No. NRF-2021R1A5A1032996).
\end{acknowledgments}

\appendix

\newcommand{\Rf}[1]{Ref.~\onlinecite{#1}}
\newcommand{\Sec}[1]{Sec.~\ref{#1}}
\newcommand{\eq}[1]{Eq.~\eqref{#1}}
\newcommand{\eqs}[1]{Eqs.~\eqref{#1}}
\newcommand{\fig}[1]{Fig.~\ref{#1}}
\newcommand{\figs}[1]{Figs.~\ref{#1}}
\setcounter{section}{0}
\setcounter{equation}{0}
\setcounter{figure}{0}
\renewcommand{\thesection}{S\arabic{section}}
\renewcommand{\theequation}{S\arabic{equation}}
\renewcommand{\thefigure}{S\arabic{figure}}

\maketitle
\section{Code availability}
Code will be made available at \url{https://github.com/park-sungjoon/topological-phase-diagram}.
We have used the GUDHI library~\cite{gudhi:urm} for TDA related computations, the Scikit-learn library~\cite{pedregosa2011scikit} for manifold learning techniques, including the mMDS and the Isomap, and the PyTorch library~\cite{paszke2019pytorch} for the minimization of the loss function.

\section{Review of TDA \label{app.tda}}
The idea of TDA~\cite{edelsbrunner2000topological,zomorodian2005computing,carlsson2009topology,edelsbrunner2010computational,chazal2017introduction} is to create a continuous shape such as the \v{C}ech complex or Vietoris-Rips complex using the data points $X$, and to use the resulting complex to infer the topology of the underlying distribution of the data points.
This requires the data points to be embedded in a Euclidean space, or a notion of distance to be defined between the data points.
In the case of wavefunctions defined over the Brillouin zone, quantum distance such as the Hilbert-Schmidt distance ($d_{HS}$) is a natural choice for the investigation of the topology of the space formed by the wavefunctions (note that in general, this is not required to be a manifold since it is possible for self-intersections to appear).
Since Vietoris-Rips complex can be created without Euclidean embedding, we use the Vietoris-Rips complex for numerical computations in this work.

Recall that the Vietoris-Rips complex $\textrm{Rips}_r(X)$, where $X$ is the data points and $r \ge 0$, is defined as the set of simplices $[x_0, ..., x_k]$ such that $d(x_i,x_j) < r$ for all pairs of $(x_i,x_j)$.
As we vary $r$, $\textrm{Rips}_r(X)$ will also evolve, and the PD is designed to capture these changes in the Vietoris-Rips complex and highlight the meaningful homological features of the geometric shape underlying the data points.
This is done by tracking how the homology groups of the simplex evolve as we change $r$, and by recording the $r$ at which the homology groups form (birth) and disappear (death).
For example, let us consider the zeroth homology group $H_0(X)$, which gives the number of connected components.
Assuming that there are $N$ data points, the simplex will consist of $N$ discrete points at $r=0$, i.e. $\{ [x_0], ..., [x_N]\}$.
Thus, for $r=0$, $H_0(X)=\mathbb{Z}^N$, since there are $N$ discrete points for (assuming that none of the points overlap exactly).
However, as we increase $r$, the distance between some of these points will become less than $r$.
Whenever this happens, the number of connected components will decrease (recall the definition of $\textrm{Rips}_r(X)$), until there is only one connected component.
Since this component will not disappear for any $r$, we indicate this by ``death'' at $\infty$.
To draw the PDs, we also similarly track the evolution of higher homology groups.
Letting $b$ and $d$ denote the $r$ at which the $k$th homology group appears and disappears, we define the PD as the set of the $(b,d)$ pairs, in addition to the diagonal line $\Delta=\{(b,d)|b=d \}$, where all the points in $\Delta$ is counted with infinite multiplicity.
The reason for introducing $\Delta$ will become clear below.

Once we obtain the PD, we have some options for the definition of the distance between them.
In this work, we chose the Wasserstein distance, but other choices such as the bottleneck distance are also possible.
To define the Wasserstein distance, let us first define a matching between diagrams $\textrm{PD}_i$ and $\textrm{PD}_j$.
A matching between $\textrm{PD}_i$ and $\textrm{PD}_j$ is defined as a subset $m\subset \textrm{PD}_i \times \textrm{PD}_j$ such that every point in $(\textrm{PD}_i-\Delta)$ and $(\textrm{PD}_j-\Delta)$ appears only once in $m$.
The Wasserstein distance $W_p(\textrm{PD}_i,\textrm{PD}_j)$ is defined as 
\begin{align}
W_p(\textrm{PD}_i,\textrm{PD}_j) = (\inf_{m} \sum_{(x,y)\in m}||x-y||_{\infty}^p )^{1/p},
\end{align}
where $|| x-y||_{\infty} $ is the $L_{\infty}$ norm in the $(b,d)$ space.
In this work, we have used $W_2$.

\section{Isometric embedding into Euclidean space \label{app.isometric}}
Although we work with the Vietoris-Rips complex, which does not rely on embedding into Euclidean space, it is still useful to know that the wavefunctions over the Brillouin zone can be embedded isometrically into the Euclidean space, which is due to the following results \cite{hioe1981n,jakobczyk2001geometry,kimura2003bloch,byrd2003characterization}.
Let $\psi$ be an $n$-component complex-valued wavefunction.
The goal is to map $\psi$ to the generalized Bloch sphere.
For this, let $\lambda_a$ be the generators of $\mathfrak{su}(n)$ algebra, which are traceless $n\times n$ Hermitian matrices.
We can choose $n^2-1$ number of generators called the generalized Pauli matrices, such that $\textrm{Tr} (\lambda_a) = 0$ and $\textrm{Tr} (\lambda_a \lambda_b) = 2 \delta_{ab}$.
Then, we can express the density matrix $\rho$ for this pure state as $\rho = |\psi\rangle \langle \psi |= \frac{1}{n} (1 + \sqrt{\frac{n(n-1)}{2}} \bm{r}\cdot \bm{\lambda})$, where $\bm{r}$ is a unit vector (generalized Bloch vector) whose components are $r_a = \frac{\langle \psi | \lambda_a | \psi \rangle}{\sqrt{2n(n-1)}}$.
Then, the Hilbert-Schmidt distance between two wavefunctions $\psi_1$ and $\psi_2$ is given by $\frac{1}{\sqrt{2}}[\textrm{Tr}((\rho_1-\rho_2)^2)]^{1/2} = \sqrt{\frac{n(n-1)}{2}}||\bm{r}_1-\bm{r}_2||$, where $||\cdot||$ is the Euclidean norm.
This shows that we can embed the wavefunction while preserving the Hilbert-Schmidt distance into the sphere $S^{n^2-2}$ with radius $\sqrt{\frac{n(n-1)}{2}}$.
Furthermore, in the case of $\mathcal{PT}$ symmetric systems, only the real symmetric generalized Pauli matrices are relevant.
Therefore, we can embed the wavefunctions into the sphere  $S^{\frac{n^2+n}{2}-2}$.

Let us note that for $n=2$ without any symmetry constraints, the Hilbert space is $\mathbb{C}P^1$ and the wavefunctions can be isometrically embedded into $S^{2}\in \mathbb{R}^3$ that has radius equal to $1/2$. 
Note that the surface area of this $S^{2}$ is $\pi$.
In the presence of the $\mathcal{PT}$ symmetry, the Hilbert space for $n=2$ is $\mathbb{R}P^1$, and the wavefunctions can be isometrically embedded into $S^{1}\in\mathbb{R}^2$ with radius $1/2$.
Note that the circumference of this $S^1$ is $\pi$.

\section{Dimensionality reduction techniques \label{app.dim_red}}
Having a good sense of what the wavefunction space looks like and how it evolves is useful, both for making sense of the PD, and for making sure that something did not go wrong during the deformation of the Hamiltonian.
For example, there should not be multiple clusters of wavefunctions after the deformation, which signals that continuity was not maintained during the deformation.
Because of the result in Appendix~\ref{app.isometric}, we could have embedded the wavefunctions directly into the Euclidean space instead of using mMDS for the SSH and the QWZ model.
However, when the data lies in higher dimension than three, dimensionality reduction techniques, including mMDS, can be useful for visualization. 
Let us note that this is another useful application of unsupervised machine learning technique in our algorithm which serves as a check that the deformation is working as intended.

\begin{figure}[t]
\centering
\includegraphics[width=8.5cm]{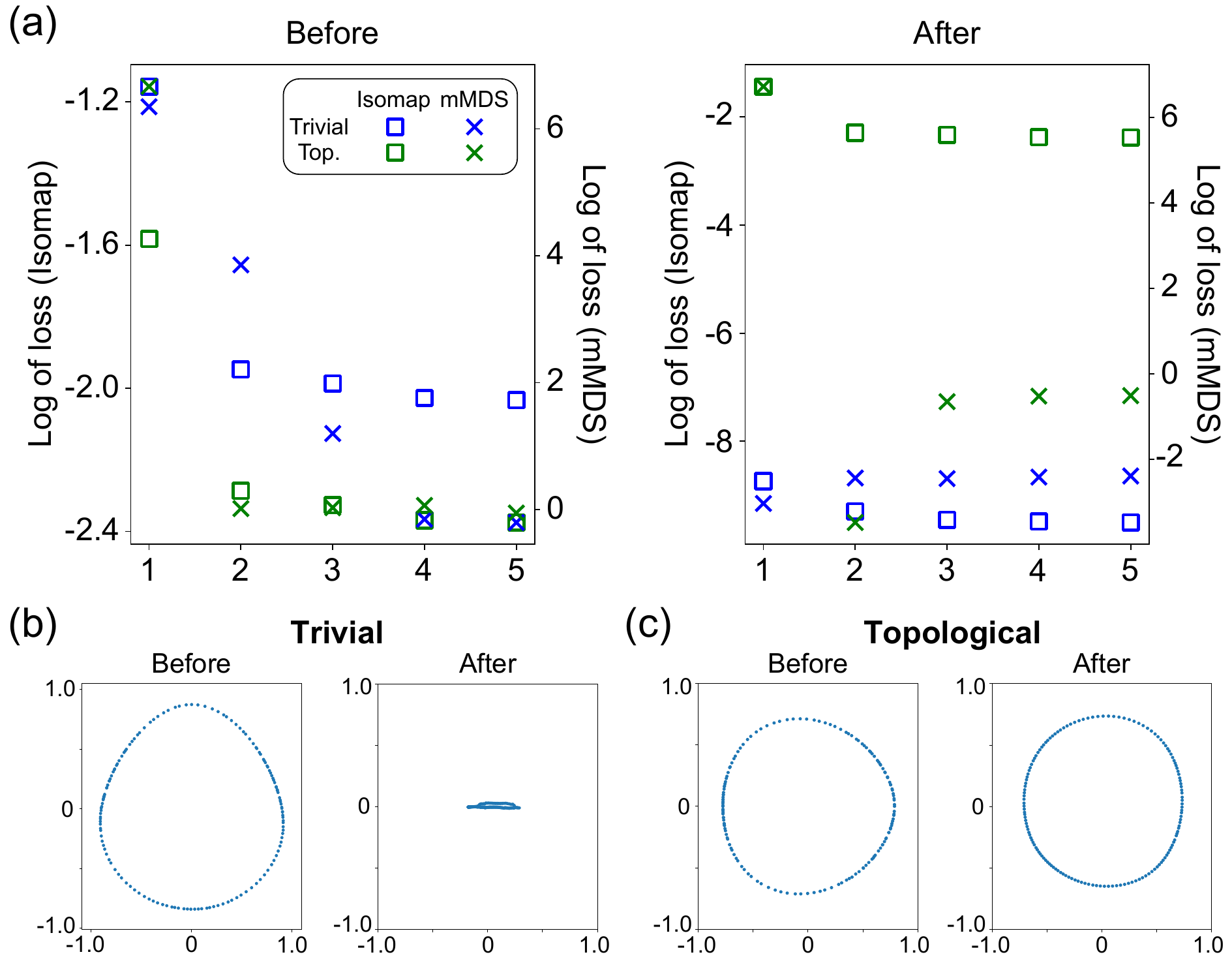}
\caption{(a) The $x$ axis is the dimension into which we embed the wavefunctions, and the $y$ axis is the logarithm of the loss resulting from Isomap (left axis) and mMDS (right axis).
The squares and $\bm{\times}$'s are for Isomap and mMDS respectively.
The blue and green colors represent topologically trivial and nontrivial phases.
The left figure is the result before the deformation, and the right figure is the result after the deformation.
(b) and (c) are results of the Isomap for topologically trivial and nontrivial states, with the same parameters as in the main text.
The left and right subfigures are results before and after the deformation.
}
\label{fig.tb_manifold}
\end{figure}

As an example, let us consider the three-band model discussed in the main text.
The result in Appendix~\ref{app.isometric} with ($n=3$) suggests that it is not possible to embed the space formed by the wavefunction into 2D or even 3D Euclidean space while preserving the Hilbert-Schmidt distance.
This can be confirmed by examining the loss values for mMDS resulting from embedding into various dimensions.
In Fig.~\ref{fig.tb_manifold}(a) (left), we show the logarithm of the loss function at various dimensions for mMDS in `$\bm{\times}$' (here, the loss is called the stress)~\cite{borg2005modern,pedregosa2011scikit} before the deformation described in the main text.
The blue $\bm{\times}$'s are the loss resulting from embedding the topologically trivial wavefunctions, and the green $\bm{\times}$'s are the loss resulting from embedding the topologically nontrivial wavefunctions.
As can be seen, the loss for the topologically trivial wavefunction cannot be embedded accurately until the dimension reaches $4$ (note that we have plotted the logarithm of the loss; embedding into two-dimension with mMDS still gives a reasonable result).

Fortunately, for the purposes of visualization, there are various other dimensionality reduction techniques we can use.
Here, we use the Isomap, whose goal is to preserve the geodesic distance~\cite{tenenbaum2000global}.
In Fig.~\ref{fig.tb_manifold}(a) (left), we show the logarithm of the loss at various dimension for Isomap in squares (here, the loss is called the reconstruction loss)~\cite{pedregosa2011scikit}.
The blue (green) squares are the loss resulting from embedding the topologically trivial (nontrivial) wavefunctions.
As can be seen, it is sufficient to embed into two dimensions for both cases, and we show the result of the Isomap in Figs.~\ref{fig.tb_manifold}(b) and \ref{fig.tb_manifold}(c) (left) for the topologically trivial and nontrivial cases, respectively.
For comparison, we also show the result of Isomap after the deformation in Figs.~\ref{fig.tb_manifold}(b) and \ref{fig.tb_manifold}(c) (right).
Also, let us observe that after the deformation, it is possible to accurately embed the wavefunctions into two-dimensional Euclidean space (Fig.~\ref{fig.tb_manifold}(a) (right)) with mMDS.

\section{Three-band model \label{app.three}}
\begin{figure}[t]
\centering
\includegraphics[width=8.5cm]{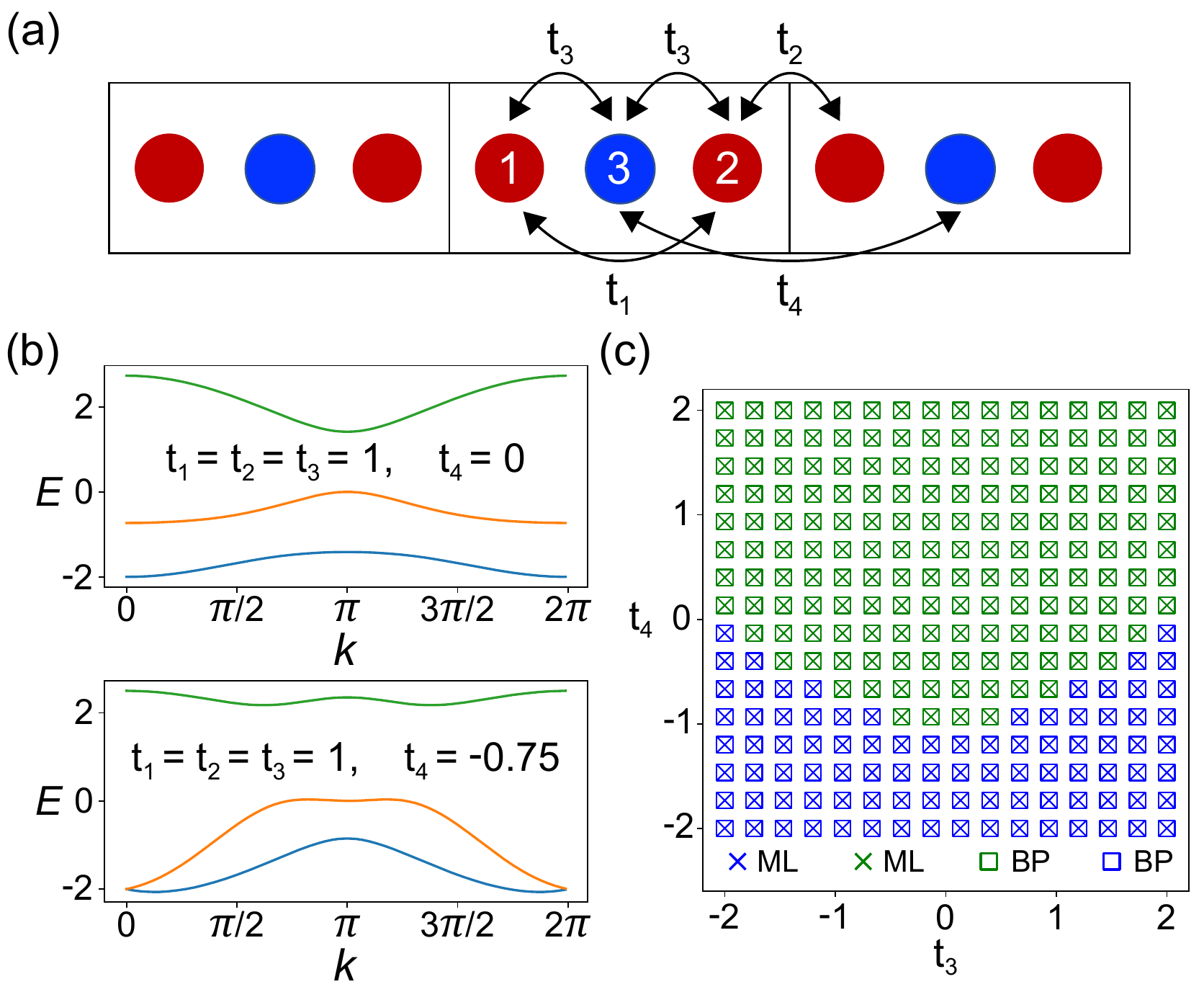}
\caption{(a) Hopping amplitudes for the three-band model. 
(b) The energy spectrum of the three-band model.
Note that when $t_1=t_2=t_3=1$, the gap closes at $t_4=-0.75$.
(c) The phase diagram shown over parameters $t_3$ (x axis), $t_4$ (y axis) ranging over $-2$ to $2$.
The `$\bm{\times}$' markers are used for the phases learned through our machine learning (ML) algorithm, and the square markers are used for phases computed by Berry phase (BP).
We indicate topologically trivial and nontrivial phase in blue and green, respectively.
}
\label{fig.tb_model_supplement}
\end{figure}

Our three-band model is created by an extension of the SSH model.
As shown in Fig.~\ref{fig.tb_model_supplement}, we have three sites per unit cell, and we have three hopping amplitudes, $t_1, t_2, t_3, t_4$.
Note that when $t_3=0$, the blue and the red atomic chains are independent of each other, and the red atomic chain forms the SSH model.
Fourier transforming the Hamiltonian, we obtain
\begin{align}
H_{\bm{k}} = \begin{pmatrix}
0 & t_1 + t_2 e^{-ik} & t_3 \\
t_1 + t_2 e^{ik} & 0 & t_3 \\
t_3 & t_3 & 2 t_4 \cos k
\end{pmatrix}.
\end{align}
Notice that this model has an inversion symmetry about the blue atom, and its representation is 
\begin{align}
\mathcal{P} = \begin{pmatrix}
0 & 1 & 0 \\
1 & 0 & 0 \\
0 & 0 & 1
\end{pmatrix},
\end{align}
so that 
\begin{align}
\mathcal{PT} = \begin{pmatrix}
0 & 1 & 0 \\
1 & 0 & 0 \\
0 & 0 & 1
\end{pmatrix}K,
\end{align}
where $K$ is the complex conjugation.
It is convenient to work in the basis in which $\mathcal{PT}=K$, which can be achieved by introducing
\begin{align}
U = \begin{pmatrix}
\frac{1+i}{2} & \frac{1-i}{2} & 0 \\
\frac{-1+i}{2} & -\frac{1+i}{2} & 0 \\
0 & 0 & 1
\end{pmatrix}.
\end{align}
Performing the change of basis $H^R_{\bm{k}}=UH_{\bm{k}}U^{-1}$ (the superscript indicating that this is in the real basis), we have
\begin{align}
H^R_{\bm{k}} = \begin{pmatrix}
t_2 \sin k & -t_1 - t_2 \cos k & t_3 \\
-t_1 - t_2 \cos k & -t_2 \sin k & -t_3 \\
t_3 & -t_3 & 2t_4 \cos k
\end{pmatrix}.
\end{align}
The energy spectrum is shown in Fig.~\ref{fig.tb_model_supplement}(b) (upper) for $t_1=t_2=t_3=1$ and $t_4=0$.
Notice that all of the bands are gapped.
In this case, the lowest energy band has $\pi$-Berry phase.
As we decrease $t_4$, the band gap closes at $k=0$ for $t_4 = -0.75$, and enters the topologically trivial phase with $0$-Berry phase.
This transition point was well captured by the phase diagram shown in the main text.
To further confirm that our method works for other parameter values, we show in Fig.~\ref{fig.tb_model_supplement}(c) the phase diagram drawn over a wider range of $t_3$ and $t_4$.
As can be seen, the classification of phases produced by our machine learning algorithm (shown in `$\bm{\times}$') is in excellent agreement with the classification based on the computation of Berry phase (shown in square)

\section{Choosing hyperparameters}
\begin{figure*}[t]
\centering
\includegraphics[width=17cm]{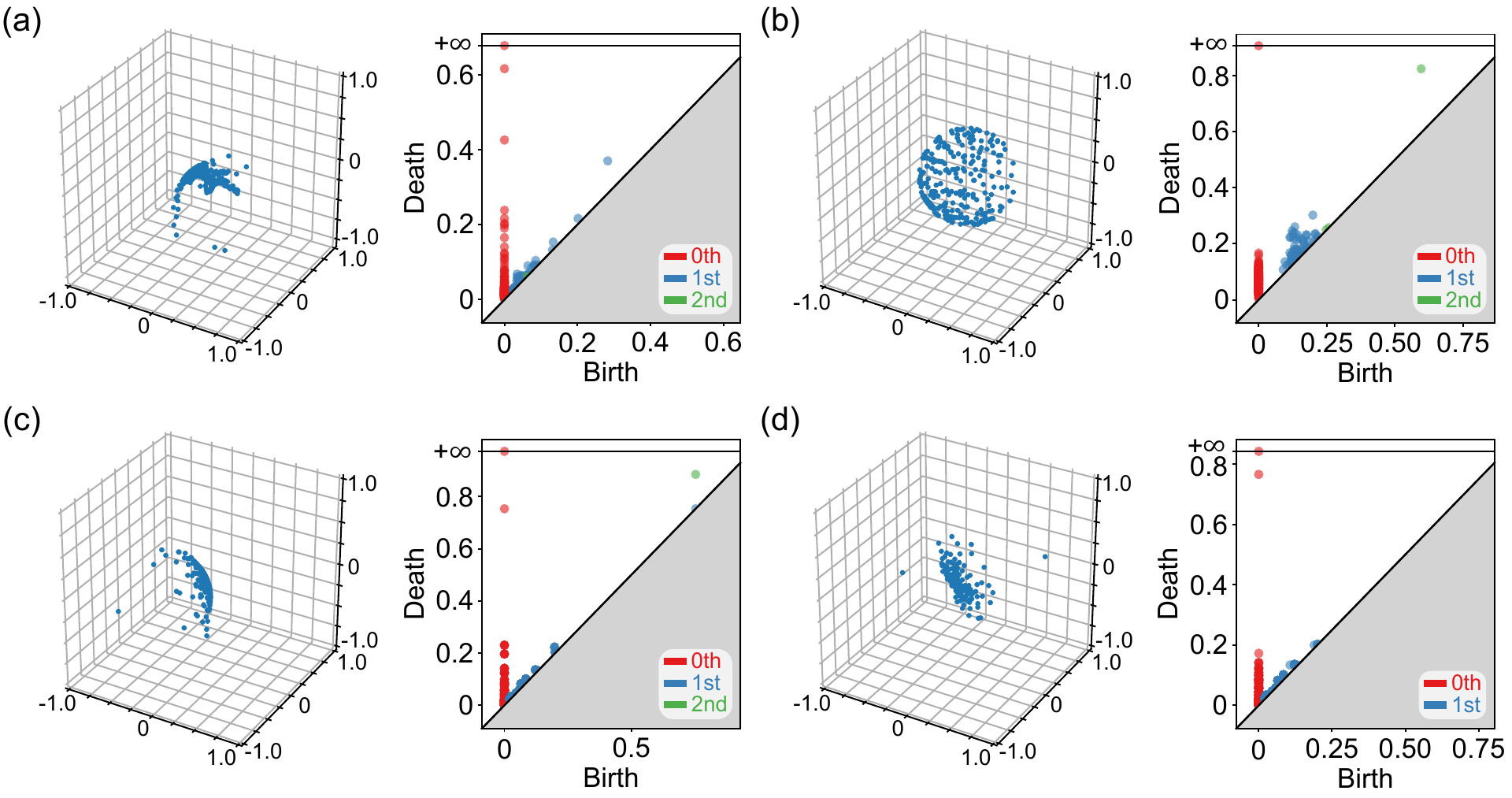}
\caption{(a) Result of mMDS (left) and the PD (right) when using a high initial learning rate ($0.01$) without giving sufficient weight to continuity loss.
(b) The case when the initial learning rate is high ($0.01$) but sufficient weight is given to the continuity loss.
(c) The case when $N_{\textrm{mesh}}=16$ is small. The results before the deformation are shown.
(d) The case when $N_{\textrm{mesh}}=16$ is small. The results after the deformation are shown.
}
\label{fig.parameter_experiment}
\end{figure*}
The important hyperparameters in our algorithm are the learning rates, and the ratios $r_s$ and $r_c$ in the loss function for the deformation.
First, let us note that we have used Adam~\cite{kingma2014adam} for minimizing Eq.~(2) in the main text.
For one-dimensional models (SSH and three-band model), we set the initial learning rate to $0.01$, and for two-dimensional model (QWZ model), we set the initial learning rate to $0.002$.

Let us next explain how we chose $r_s$ and $r_c$.
For $r_s$, we begin by setting $r_s = L_v(\Theta^{e=0}) /L_s(\Theta^{e=0}; r_s = 1)$, where $e$ is the epoch (indexed from $0$).
Then, we update $r_s = L_v(\Theta^{e}) /L_s(\Theta^{e}; r_s = 1)$ every $10$ epochs.
This choice of $r_s$ is due to the observation that as the number of points in the $k$ mesh increases, the sum of the volume tends to a constant (total volume), but the squared sum of the volume decreases.
This update rule is imposed to give similar weight to the two objectives.
For $r_c$, we update $r_c = L_v(\Theta^{e}) /L_c(\Theta^{e}; r_c = 1)$ while $e$ is less than some fixed integer, which we chose to be $10$ for 1D models and $50$ for 2D models.
This is to ensure that initially, the deformation starts in the direction that maintains the continuity.
Then, we slowly anneal $r_c$ to $1$ by multiplying $r_c$ by a fixed ratio (for 1D models, we do this over 100 epochs and for QWZ, we do this over 500 epochs since we are using smaller learning rate).
After the annealing process, we make sure that $L_c$ does not become too large by 
keeping $L_c < 0.5 L_v$.

We next show some experiments we have done that led us to the above procedure.
Let us first discuss our choice of the learning rates.
The reason for setting the learning rate for two-dimensional model lower than for one-dimensional model is that the deformation of the wavefunctions need to be more coordinated with each other, since an increase in the dimensionality results in an increase in the coordination number for the wavefunctions sampled from the mesh in the momentum space.
If we do not use the update rule for $r_s$ and $r_c$ in the loss function but use the learning rate of $0.01$, there are situations in which the continuity of the wavefunction is not kept during the deformation.
We believe that this happens because when the Hilbert-Schmidt distance between two wavefunctions is large, more freedom is allowed for the wavefunction deformation, so that when the wavefunctions are initially deformed such that the continuity is not maintained, the following deformation will destroy the continuity altogether.
We illustrate this in Fig.~\ref{fig.parameter_experiment}(a). 
As can be seen, the wavefunctions of the topological phase no longer forms a sphere after the deformation, and the second homology group does not even appear in the PD.
Using the update rule below for the ratios in the loss function fixes this problem, which supports our hypothesis that above.
We show this in Fig.~\ref{fig.parameter_experiment}(b), where it can be seen that even if we use the initial learning rate of $0.01$, the wavefunction still forms a sphere after the deformation.
This result indicates that it is possible to use a higher initial learning rate.
However, but we maintain smaller initial learning rate of $0.002$ for the 2D problem for safety purposes.

Finally, let us discuss the choice of $N_{\textrm{mesh}}$.
For the SSH model, we used $N_{\textrm{mesh}}=100$ and for the three-band model, we used $N_{\textrm{mesh}}=150$.
For the QWZ model, we used $N_{\textrm{mesh}}=40$, so that in total, $1600$ points are sampled from the Brillouin zone.
However, after the deformation, we sparsify points by trimming away wavefunctions that are too close to each other, until only $240$ points are left.
To illustrate the importance of sufficiently sampling points in the momentum space, let us consider the QWZ model, and show the result for $N_{\textrm{mesh}}=16$ while keeping everything else the same (i.e. learning rate is set to $0.002$ and we use the update rule as above).
Before the deformation, we see that the sampled points form a poor representation of the sphere, see Fig.~\ref{fig.parameter_experiment}(c).
For this reason, the deformation does not behave as intended, as shown in Fig.~\ref{fig.parameter_experiment}(d).
Note that the second homology does not even appear in the PD after the deformation.
Such behavior occurs when the band gap between the occupied and the unoccupied bands become small, so that it is important that we use a large $N_{\textrm{mesh}}$ and trim away points before computing the PD.

\section{Minima of volume \label{app.min_vol}} 
An advantage of our method is the interpretability of the objective of minimizing the volume (1D: length, 2D: area) of the curve or surface formed by the wavefunction.
For 1D, we are essentially looking for closed curves that are also geodesics, and for 2D and higher, we are looking for closed surfaces that are minimal.
Since we are discretizing the problem and minimizing the loss function in Eq.~(2) in the main text, what we are actually doing does not exactly match with the above objective.
Nevertheless, the two objectives are similar, and it is of interest to know whether there is a mathematical justification of the objective of finding curves and surfaces with minimum volume.

Let us first consider the 1D case for systems with the $\mathcal{PT}=K$ symmetry, so that the wavefunctions lie in the space $\mathbb{R}P^{n}$, where the total number of energy bands is $(n+1)$. 
If we assume that the wavefunctions form a smooth curve, we can make some guarantees on the objective as follows.
Let us first consider for simplicity the smooth curves in $\mathbb{R}P^{1}$.
It is clear from the existence and uniqueness of geodesics that a closed curve with minimal length that is not a constant is just the arc that connects two antipodal points of $S^{1}$.
Note that here, instead of isometrically embedding $\mathbb{R}P^n$ in the Euclidean space, we are considering $\mathbb{R}P^{n}$ as a unit sphere $S^n$ with the antipodal points identified, such that the distance between two points is defined by $d_{HS}$.
Such a closed geodesic is also a generator of the fundamental group of $\mathbb{R}P^{1}$, so that we can obtain a one-to-one correspondence between closed geodesics and the fundamental group of $\mathbb{R}P^{1}$.
It is interesting to note that geodesics have length $|n|\pi$, where $n$ is the integer representing the element of the $\mathbb{Z}$-valued fundamental group of $\mathbb{R}P^{1}$.
Therefore, assuming that we have reached the minimum of the length in the closed curve in the class of its fundamental group, it is possible to distinguish between topological insulators through the orientation of the geodesic and its length.

For $\mathbb{R}P^{n}$ with $n\ge 2$, it is clear from the homogeneity and the isotropicity of $\mathbb{R}P^{n}$ that the closed geodesics are essentially the same as those in $\mathbb{R}P^{1}$ (i.e. they lie in $\mathbb{R}P^{1}\subset $ $\mathbb{R}P^{n}$).
Since the first homotopy group of $\mathbb{R}P^{n}$ with $n\ge 2$ is $\mathbb{Z}_2$, one may at first expect that the algorithm will not work because of the existence of geodesics with `$2n$ windings' within $\mathbb{R}P^{1}\subset \mathbb{R}P^{n}$.
However, such geodesics are not stable in the sense that there are small perturbations that will decrease its length.
To see this, it suffices to consider $\mathbb{R}P^2$, since we can canonically embed $\mathbb{R}P^2$ in $\mathbb{R}P^n$ for $n>2$.
Note that we can parametrize the wavefunctions in $\mathbb{R}P^2$ as
\begin{align}
\psi = \begin{pmatrix}
x \\
y \\
\sqrt{1-x^2-y^2}
\end{pmatrix}.
\end{align}
Then, an infinitesimal length is given by $ds^2 = g_{\mu \nu} dx^{\mu} dx^{\nu}$, where $dx^\mu = (dx,dy)$, and 
\begin{align}
g_{\mu \nu} = \frac{1}{1-x^2-y^2}\begin{pmatrix}
1-y^2 & xy \\
xy & 1-x^2
\end{pmatrix}.
\end{align}
Changing the variables to $(r,\theta)$ where $x=r\cos \theta$ and $y=r \sin \theta$, we have 
\begin{align}
ds^2 = \frac{1}{1-r^2}dr^2 + r^2 d\theta^2.
\end{align}
Now, let us consider circles with fixed $r$, so that $ds^2 = r^2 d\theta^2$.
Since the circle in $S^2$ with $r=1$ is the geodesic that winds twice in the $\mathbb{R}P^1\subset \mathbb{R}P^2$ (note that $\mathbb{R}P^2$ is the $S^2$ with antipodal points identified) , small deformations that reduces $r$ reduce the length of this circle, and we see that such a geodesic is not stable.
Combining everything, we see that there is a one-to-one correspondence between closed geodesics that are stable and the fundamental group of $\mathbb{R}P^{n}$, for $n\geq 1$.

Next, let us consider the 2D case without symmetry constraints, so that the wavefunctions lie in the space $\mathbb{C}P^{n}$.
This case is much more complicated, but we can still make some guarantees when the surface formed by the wavefunctions forms a smooth submanifold of  $\mathbb{C}P^{n}$.
From Ref.~\cite{lawson1973stable}, a closed integral current in $\mathbb{C}P^n$ is stable (and minimal) if and only if it is an integral chain of algebraic variety.
Also, a projective algebraic variety is a complex manifold if it is smooth \cite{ballmann2006lectures}, i.e. has no singularities.
Finally, a corollary of the Wirtinger inequality is that every complex submanifold of a K{\"a}hler manifold is volume minimizing in its homology class \cite{federer2014geometric,harvey1982calibrated}.
Combining these, it follows that if a smooth submanifolds of $\mathbb{C}P^n$ is minimal and stable, it is a smooth complex projective variety, i.e. smooth complex submanifold of  $\mathbb{C}P^n$.
Therefore, it is also volume minimizing in its homology class.
Thus, if the wavefunctions form a smooth submanifold of $\mathbb{C}P^n$, it will at least not get stuck in a local minimum of the area functional.

It is also interesting to note that since $\mathbb{C}P^1$ (a sphere with area $\pi$) is a complex submanifold of $\mathbb{C}P^n$,  it is volume minimizing in its homology class (note that $H_{0}(S^2)=\mathbb{Z}$, $H_{1}(S^2)=0$, $H_{2}(S^2)=\mathbb{Z}$).
Since $\mathbb{C}P^n$ is simply connected, and the homology groups of the torus over $\mathbb{Z} $ are 
$H_{0}(T^2)=\mathbb{Z}$, 
$H_{1}(T^2)=\mathbb{Z} \oplus \mathbb{Z}$, 
$H_{2}(T^2)=\mathbb{Z}$, 
$H_{n>2}(T^2)=0$, 
the homology class of continuous maps from $T^2$ to $\mathbb{C}P^n$ is determined by the second homology.
We should therefore expect that if we deform the surface formed by the wavefunction, reach a stable minimum, and if the resulting surface is smooth, it will have area $\pi |n|$, where $n\in \mathbb{Z}$ is the second homology class of the surface in $\mathbb{C}P^n$, see also Refs.~\cite{roy2014band,peotta2015superfluidity,mera2021relating,mera2021k,zhang2021revealing,ozawa2021relations}.

\section{Oriented area indicator of Chern number \label{app.oriented_area}}
\begin{figure}[t]
\centering
\includegraphics[width=8.5cm]{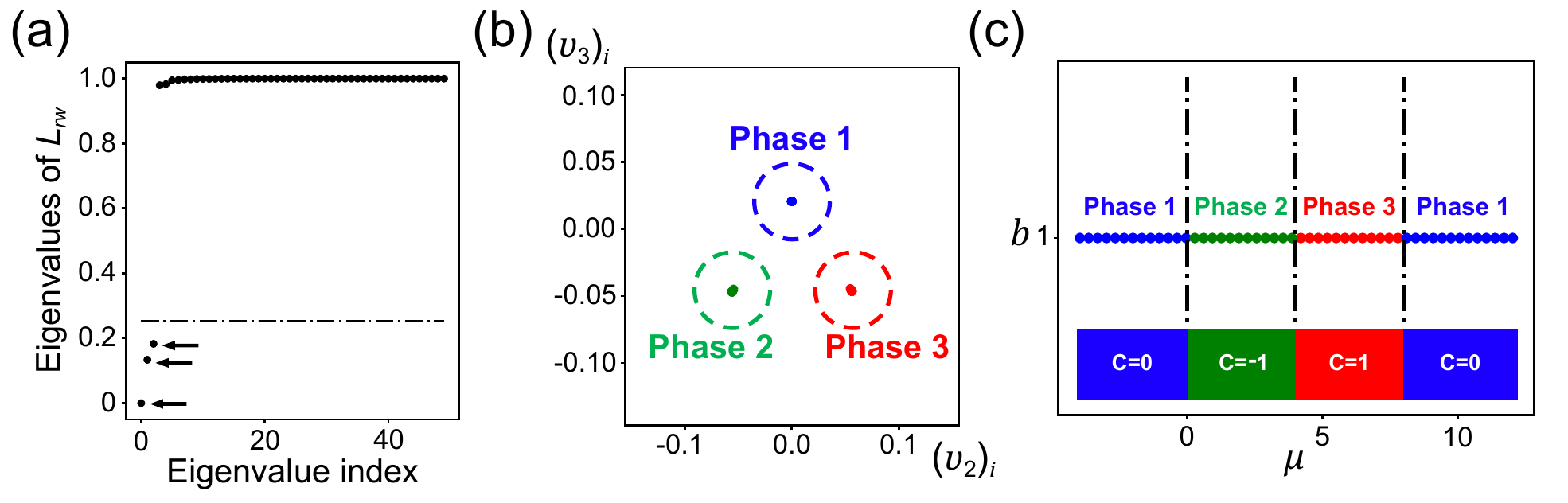}
\caption{
(a) Eigenvalues of random walk Laplacian computed using the oriented area. (b) From the plot of $v_2$ and $v_3$, we can see three clusters.  
(c) Using the three clusters, we can plot the phase diagram. 
Note that the phase boundary at $\mu=4$ between Chern number $\pm 1$ is correctly captured.
}
\label{fig.QWZ_oriented_area}
\end{figure}
As mentioned in the main text in the section ``QWZ model'', the oriented area can faithfully capture the Chern number.
This is because after the area minimization, the Chern number is determined by the map $T^2\rightarrow S^2$ as discussed in Appendix~\ref{app.min_vol}. 
Specifically, the number of times the map winds around $S^2$ determines the size of the Chern number, and the orientation of the winding determines the sign of the Chern number.
Both the area swept by the wavefunction and the orientation can be determined using the oriented area.
Let us first explain the general procedure and apply it to the QWZ model.

First, we proceed as in the main text and minimize the area of the map $T^2 \rightarrow \mathbb{C}P^{n-1}$, where $n$ is the total number of bands. 
This will deform the map to $T^2 \rightarrow \mathbb{C}P^{1}\approx S^2 \in \mathbb{R}^{n^2-1}$ (recall that we can isometrically embed the wavefunctions in to the Euclidean space, see Appendix~\ref{app.isometric}).
Second, we find the 3D plane in which this deformed map lives through principle component analysis.
Let $\hat{x},\hat{y},\hat{z}$ be the orthonormal basis for this 3D plane.
Third, working with this fixed basis, we can compute the oriented area, as will be explained below.
Finally, we can cluster phases with similar oriented area.

Now, let us define the oriented area.
Let $\bm{V}_{\bm{k}}$ be the map $T^2 \rightarrow S^2$ obtained by the area minimization, expressed in the $\hat{x},\hat{y},\hat{z}$ coordinates.
First, we discretize $T^2$ into $(k_{x,i},k_{y,j})$, where $k_{x,i},k_{y,j} \in \{2\pi n/N_{\textrm{mesh}} | n=0,…,N_{\textrm{mesh}}-1\}$. 
Because we have defined the 3D coordinate system, $\hat{x}, \hat{y}, \hat{z}$, we can work with the familiar 3D Euclidean space. 
Using the fact that cross product gives the oriented area, we can compute the oriented area as follows:
$A[\bm{V}_{\bm{k}}] = \sum_{ij} \hat{\bm{V}}_{k_{x,i},k_{y,j}} \cdot [(\bm{V}_{k_{x,i+1},k_{y,j}} - \bm{V}_{k_{x,i},k_{y,j}})\times(\bm{V}_{k_{x,i},k_{y,j+1}} - \bm{V}_{k_{x,i},k_{y,j}})]$.

To compare the oriented area for $\bm{V}_{\bm{k}}^{1}$ and $\bm{V}_{\bm{k}}^{2}$ resulting from two different parameters, we bring the two surfaces into the same 3D plane, so that there is no ambiguity in the orientation of the coordinate system.
This can be achieved by considering the wavefunctions $|\psi_{\bm{k}}^{1} \rangle$ and $|\psi_{\bm{k}}^{2} \rangle$ as point cloud in $\mathbb{R}^{n^2-1}$ through the map $\langle \psi^{1,2}_{\bm{k}_\alpha} | \lambda_a | \psi^{1,2}_{\bm{k}_\alpha} \rangle$ discussed in Appendix~\ref{app.isometric}, where $\alpha = 1, ..., N_{\textrm{mesh}}^2$.
We can then bring the point cloud formed by $|\psi^1_{\bm{k}_\alpha}\rangle$ to the point cloud formed by $|\psi^2_{\bm{k}_\alpha} \rangle$ through the transformation $|\psi_{\bm{k}_\alpha}^1\rangle \rightarrow U(\bm{\theta}) | \psi_{\bm{k}_\alpha}^1\rangle$, where $U(\bm{\theta})\in SU(n)$ and $\bm{\theta}$ parameterizes the $SU(n)$ matrix.
Note that $U(\bm{\theta})$ does not depend on $\alpha$.
The above problem is a variant of the point cloud registration.
Let us now explain how $U(\bm{\theta})$ can be found. 
First, for each $| \psi_{\bm{k}_\alpha}^1\rangle$, let $| \psi_{\bm{k}_\alpha}^2\rangle$ be the wavefunction that minimizes the distance to $|\psi_{\bm{k}_\alpha}^1\rangle$.
Second, find $\bm{\theta}$ that minimize $\sum_\alpha d^2_{HS}(U(\bm{\theta})| \psi_{\bm{k}_\alpha}^1\rangle, | \psi_{\bm{k}_\alpha}^2\rangle)$. 
Third, repeat the above two steps until convergence.
Once we find $U(\bm{\theta})$, we can compute the oriented area in the coordinates $\hat{x},\hat{y},\hat{z}$ found for the union of the point clouds $U(\bm{\theta})|\psi_{\bm{k}_\alpha}^1\rangle$ and $|\psi_{\bm{k}_\alpha}^2\rangle$

For the QWZ model, we can skip the point cloud registration step because the wavefunctions already live in $\mathbb{R}^3$. 
We can therefore directly compute the oriented area and cluster different topological phases by replacing the Wasserstein distance between persistence diagrams (i.e. $W_2 (\textrm{PD}_i, \textrm{PD}_j)$) with difference in oriented area $|A[\bm{V}^i_{\bm{k}}]-A[\bm{V}^j_{\bm{k}}]|$ in the process of spectral clustering.
From the eigenvalues of the random-walk Laplacian $L_{\textrm{rw}}$, we see that there are three clusters, as shown in Fig.~\ref{fig.QWZ_oriented_area} (a), as is clear from the plot of the components of the eigenvectors, $v_2$ and $v_3$, in Fig.~\ref{fig.QWZ_oriented_area}. 
Let us note that the component of the first eigenvector $v_1$ are the same (i.e. $(v_1)_{i}=(v_1)_j$ for all $i,j$), and is omitted in the plot for simplicity.
As can be seen from Fig.~\ref{fig.QWZ_oriented_area}, the phase diagram of the QWZ model can faithfully be captured.

\begin{figure}[t]
\centering
\includegraphics[width=8.5cm]{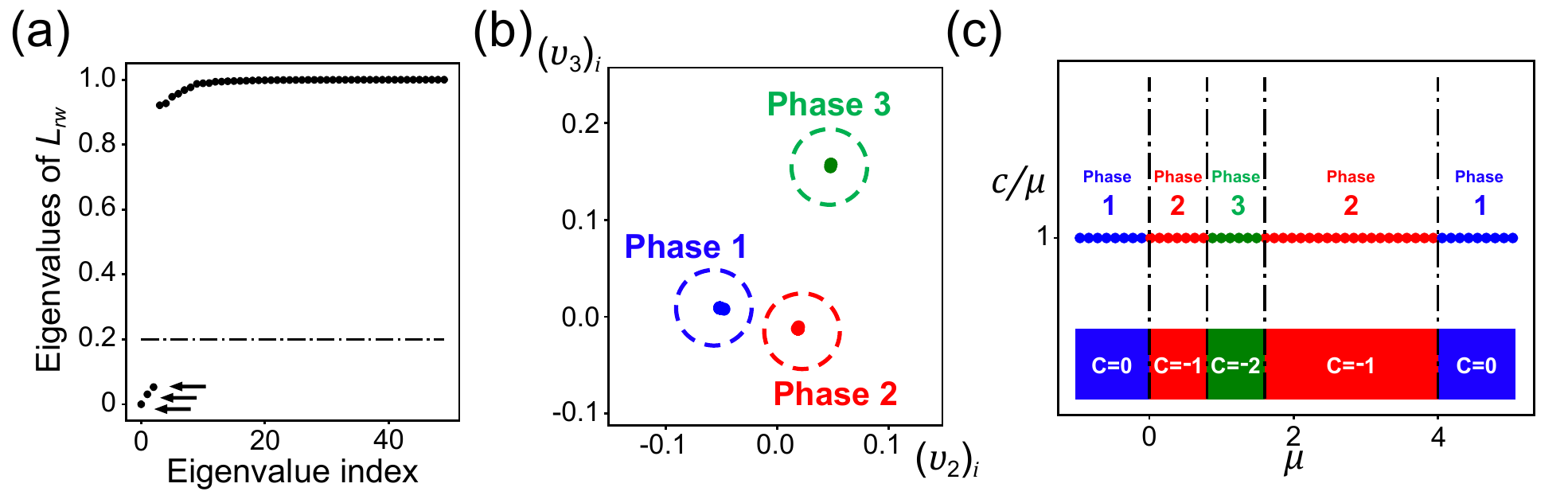}
\caption{
(a) Eigenvalues of random walk Laplacian computed using the oriented area. (b) From the plot of $v_2$ and $v_3$, we can see three clusters.  
(c) Using the three clusters, we can plot the phase diagram. 
All of the different Chern numbers are correctly captured.
Note that the phase boundaries are indicated in black vertical dashed lines at $\mu=0, 0.8, 1.6, 4$. 
}
\label{fig.mod_QWZ_oriented_area}
\end{figure}

Because the QWZ model does not show Chern numbers of magnitude greater than 1, we also study a modified QWZ model: 
$[(1-c)  \sin k_x +c \sin 2k_x] \sigma_x + \sin k_y  \sigma_y + [\mu-2(2-(1-c)  \cos k_x - c \cos 2k_x -\cos k_y )] \sigma_z$.
For $c=0$, this is just the QWZ model with $b=1$, and for $c=1$, this is the QWZ model with $b=1$ with the replacement $k_x \rightarrow 2k_x$. 
For simplicity, we study this model with $\mu =c$.
For $\mu < 0,C=0$; $0<\mu<0.8, C=-1$; $0.8 < \mu < 1.6$, $C=-2$; $1.6 < \mu < 4, C=-1$; $\mu>4, C=0$
Note that the oriented area method correctly distinguishes these phases, as can be seen in \fig{fig.mod_QWZ_oriented_area}.

On the other hand, it is not immediately clear whether the oriented area can be generalized to arbitrary symmetry class.
For example, in $PT$ symmetric 1D system, if the total number of energy bands is $2$, the topological invariant is $\mathbb{Z}$.
This can be captured with the oriented length defined similarly to the oriented area.
However, when the number of energy bands is greater than $2$, the topological invariant is $\mathbb{Z}_2$.
In this case, the orientation loses its meaning and only the absolute size of the minimum length is relevant.
Thus, volume based methods can require further analysis to be useful, whereas the TDA based method used in the main text can easily be applied without detailed knowledge about the topological invariant for the symmetry class. 
This is the main reason for adopting the TDA based method instead of the volume based method in the main text.

\bibliography{reference.bib}

\end{document}